\documentclass{pasj00}  

\begin{document}
\SetRunningHead{S. Oikawa and Y. Sofue}{Rotation Curve Anomaly and Warping of M51}
\Received{}  \Accepted{2014} 

\def\kms{km s$^{-1}$}  \def\Msun{M_\odot} \def\be{\begin{equation}}
\def\ee{\end{equation}} \def\bc{\begin{center}} \def\ec{\end{center}} 
\def\r{\bibitem[]{}} \def\vrot{V_{\rm rot}}  \def\Vrot{V_{\rm rot}}
\def\vbul{V_{\rm bul}} \def\vdisk{V_{\rm disk}} \def\vhalo{V_{\rm halo}} 
\def\deg{^\circ} \def\vr{v_{\rm r}}

\title{Rotation Curve Anomaly and Galactic Warp in M51}
\author{Shouta {\sc Oikawa}$^{1,\dagger}$ and Yoshiaki {\sc Sofue}$^{1,2,\ddagger}$}  
\footnote{Present address on leave from Meisei University: $\dagger$ Ministry of Defence, Shibata, 957-8530  Niigata; $\ddagger$ The University of Tokyo.}
\affil{1. Department of Physics, Meisei University, Hino-shi, 191-0052 Tokyo \\
2. Institute of Astronomy, The University of Tokyo, Mitaka, 181-0015 Tokyo, \\ 
Email:{\it sofue@ioa.s.u-tokyo.ac.jp}}
\maketitle
\begin{abstract} 
We revisit the anomaly of rotation curve in the nearly face-on galaxy M51 that shows an apparently faster decrease of rotation velocity than the Keplerian law in the outer disk, further showing apparent counter rotation in the outermost HI disk. We interpret this anomaly as due to warping of the galactic disk, and determined the warping structure of M51's disk using the tilted-ring method, assuming that the intrinsic rotation curve is normal. It is shown that the disk is nearly flat in the inner disk at a constant inclination angle, but the disk suddenly bends at radius 7.5 kpc by about 27$\deg$. The inclination angle, then, decreases monotonically outward reaching a perfect face-on ring at 18 kpc, beyond which the disk is warped in the opposite sense to the inner disk, resulting in apparent counter rotation. 
\end{abstract}

\section{Introduction}
Spiral galaxies have universally flat rotation curves (Rubin et al. 1980; Persic and Salucci 1996; Salucci et al. 2001; Sofue and Rubin 2001). However, two exceptional cases of anomalously rapid decrease in rotation velocity have been known: one in the edge-on peculiar galaxy M82 (NGC 3032) and the other in the face-on Sc spiral M51 (NGC 5194). 

It was shown that the rotation curve of M82 is fitted by the Keplerian law at radii beyond $\sim 3$ kpc. The Kepler rotation was interpreted as due to absence of dark halo by tidal truncation during the past gravitational encounter with the parent galaxy M81 (Sofue 1998). For an edge-on galaxy like M82, the observed radial velocity can be almost directly converted to rotation velocity for negligible correction of inclination, representing the real kinematics of the galactic disk.

On the other hand, rotation curve for a face-on galaxy is sensitive to the inclination angle. The face-on galaxy M51 has flat rotation in the inner disk, but the curve suddenly bends at radius $2'.5$ (7.5 kpc) kpc, and decreases faster than the Keplerian law (Sofue et al. 1996).
M51's rotation curve has been obtained at various wavelengths to exhibit high-accuracy in optical (Tully 1974), HI (Roberts and Warran 1970; Haynes et al. 1978; Tilanus and Allen 1990; Rots et al. 1990; Rand et al. 1993), and CO line observations (Garcia-Burillo et al.1993; Nakai et al. 1994; Kuno et al. 1995; Kuno and Nakai 1997; Koda et al. 2001; Shetty et al. 2007). Observations showed that the CO-line rotation curve in the molecular gas disk is nearly flat, whereas HI-line observations showed apparently decreasing velocity beyond $\sim 8$ kpc. Even counter rotation was observed in outermost HI disk (Appleton et al. 1987; Rots et al. 1990). 

In this short note, we revisit the anomaly of apparent rotation curve of M51, and interpret it as due to warping of the disk. 

\section{Anomalous Rotation Curve in M51}

\subsection{Apparent bend of rotation curve at 7.5 kpc}

\begin{figure}
\begin{center} 
\includegraphics[width=8cm]{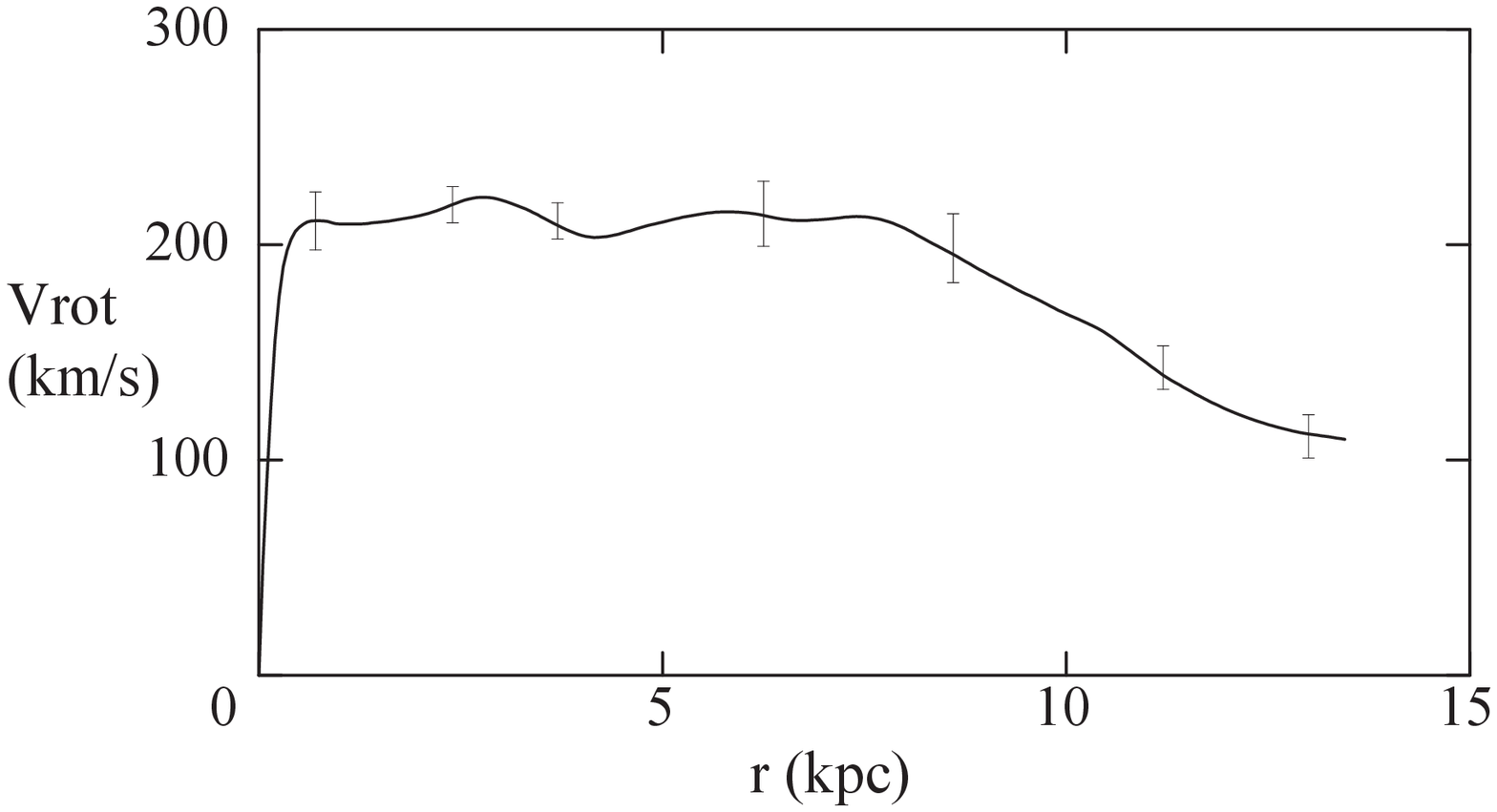}     
\end{center}
\caption{Observed rotation curve of M51 obtained for a constant inclination $i=24\deg$ using the data from Sofue (1996, 1997), who used $i=20\deg$. Error bars are indicated at representative radii.  } 
\label{rcM51}  

\begin{center} 
\includegraphics[width=7cm]{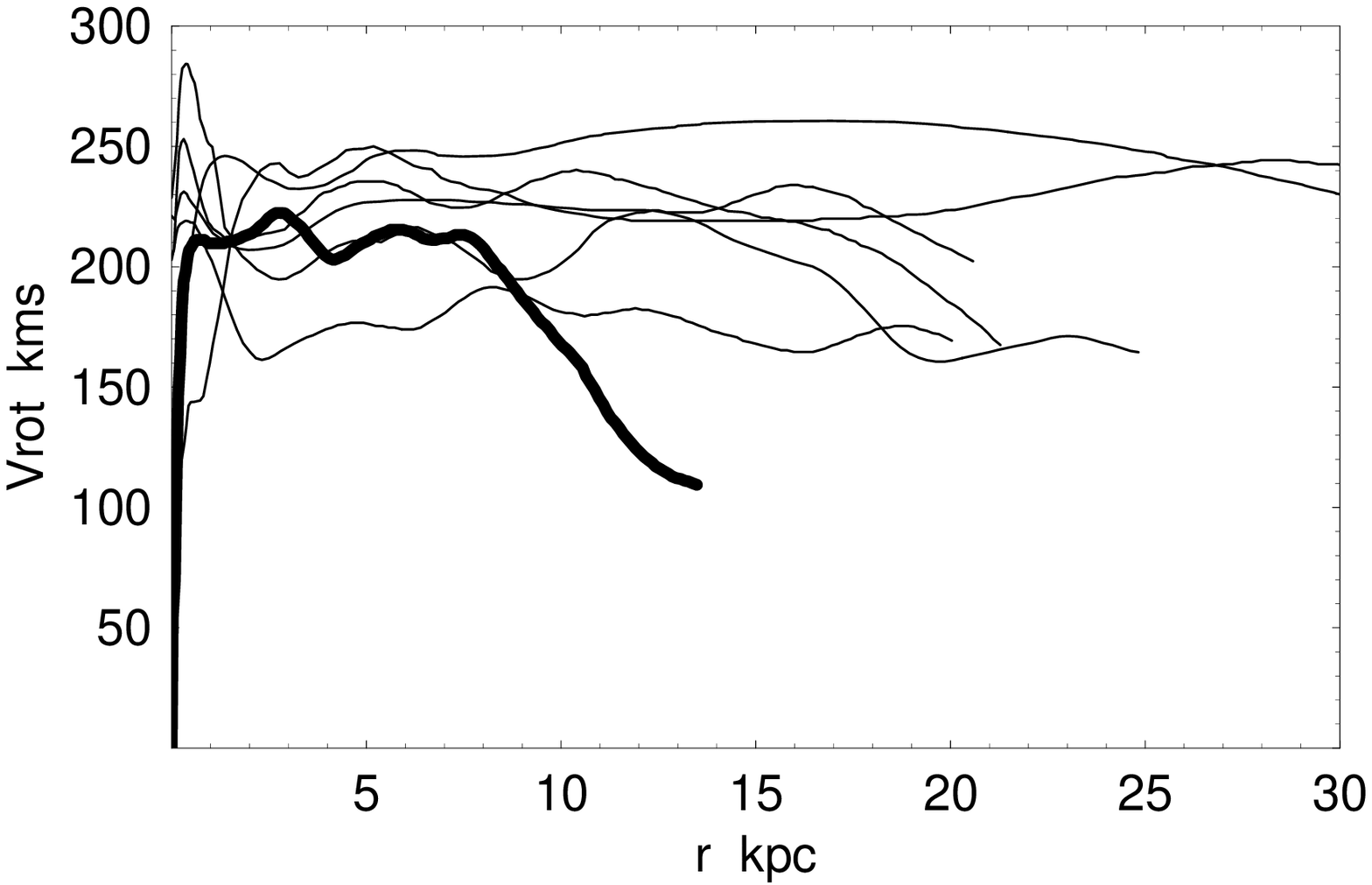}   
\end{center}
\caption{Rotation curve of M51 (thick line) compared with rotation curves of the Milky Way, M31, NGC 891, and NGC 3079 by thin lines (Sofue et al. 1999).} 
\label{rcobs}  
\end{figure}

Figure \ref{rcM51} shows a rotation curve of M51 obtained by Sofue (1996) from compilation of observations in the H$\alpha$, CO and HI line emissions. The original curve in Sofue (1996) was calculated for an inclination $i = 20\deg$, while the curve here has been re-calculated using a more recent inclination value of the inner main disk, $ i = 24\deg$ (Shetty et al. 2007).
The rotation curve is nearly flat in the inner disk at $r < 7$ kpc. However, it  bends suddenly at $r=7.5$ kpc, beyond which the velocity decreases faster than the Keplerian law. In figure \ref{rcobs} we compare M51's curve with those of typical disk galaxies, which exhibit nearly flat rotation until their edges.

\subsection{Counter rotation in the outermost HI disk}

The decreasing rotation velocities at larger radii are clearly observed in HI-line velocity fields (Rots et al. 1990; Tillanus and Allen 1990). 
Using the HI velocity field presented by Rots et al. (1990), we read the contour values of radial velocities and corresponding radii along the major axis fixed at a position angle of $\phi=162\deg$. Thereby, we did not use northern data showing systemic velocities higher than 600 \kms around the companion galaxy NGC 5195, where HI gas is supposed to be strongly disturbed, except for one point at $r=10'$ with 580 \kms. Hence, the northern data are less accurate, while consistent with the southern measurements.

\begin{figure}
\begin{center} 
\includegraphics[width=7cm]{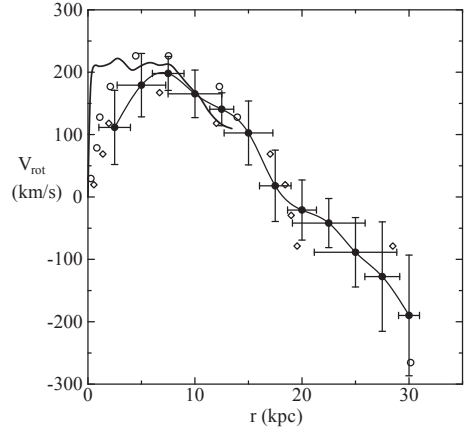}   
\end{center}
\caption{Apparent HI rotation curve obtained from the contour values in the velocity field by Rots et al (1990: their Fig. 7) for a fixed inclination $i=24\deg$and position angle $\phi=162\deg$. The read data are shown by openl circles and diamonds for the northern and southern halves, respectively.  Filled circles with error bars are Gaussian-running-averaged rotation velocities with averaging half width 2.5 kpc calculated every 2.5 kpc radius interval. Inserted inner thick line is the rotation curve from figure \ref{rcM51}.} 
\label{rotsAv} 
\end{figure}

The measured velocities are shown by small circles (northern half) and triangles (southern half) in figure \ref{rotsAv}. The values are differences of radial velocities and the systemic velocity of $V_{\rm sys}=470$ \kms, and corrected for the assumed constant inclination angle of $i=24\deg$. Rotation velocities are plotted after mirror-rotating superposition of the northern and southern measurements. 

The measured values are, then, averaged by Gaussian-weighted running mean with a half width of 2.5 kpc at every 2.5 kpc radius interval. The obtained rotation velocities are plotted by large dots with error bars in figure \ref{rotsAv}. HI velocities at $r <\sim 7$ kpc were not used in the analysis, because of missing HI gas in the inner region. The number of read data points (contour values) beyond $r\sim 20$ kpc are only one in the northern half and two in the south, so that the fitted rotation curve at $r>20$ kpc has larger uncertainty compared to that within 20 kpc.

HI rotation curve from $r\sim 7$ kpc to 13 kpc show a good agreement with those in figure \ref{rcM51}. The apparent rotation velocity decreases monotonically upto $r\sim 30$ kpc. It becomes nearly zero at $r\sim 18$ kpc, and further decreases to $\sim -200$ \kms at the measured edge. The bend and monotonical decrease of rotation curve are observed systematically both in the northern and southern disks. This implies that the anomaly may not be due to local velocity perturbations, but can be attributed to general kinematics of the whole galactic disk.

\section{Tilted-Ring Method}

The rotation velocity $\vrot$, radial velocity $\vr$, and  inclination angle $i$ in a galactic disk are coupled to each other by
\begin{equation}
\vr(r,\theta) =V_{\rm obs}(r,\theta)- V_{\rm sys}= \vrot(r) \cos \theta \sin i,
\end{equation}
where $\theta$ is azimuth angle in the disk of a measured point from the major axis, $V_{\rm obs}(r,\theta) $ is measured radial velocity and $V_{\rm sys}$ is systemic velocity of the galaxy. The position angle and azimuth angle are related by
\be
\theta(\phi)={\rm atan} \left({{\rm tan} \phi / {\cos}~i} \right).
\ee 

\subsection{Simultaneous determination of $i$ and $\vrot$} 

If a velocity field is observed, coupling of rotation velocity and inclination can be solved using the tilted-ring technique  (Rogstad et al. 1974; Bosma 1981; Begeman 1987; J{\'o}zsa et al. 2007).  This is due to the functional shape of variation of $\vr(r,\theta)/\vr(r,0)$ against the position angle on the sky $\phi$, which is uniquely related to the inclination angle $i$ and azimuth angle $\theta$. Here, $\vr(r,0)$ is the maximum value of $\vr$ along an initially chosen ring. The inclination angle $i$ is determined iteratively by comparing observed and calculated $\vr$ variations with $\phi$. Once $i$, and simultaneously $\vrot=\vr(r,0)/{\rm sin} ~i$, are determined, the same process is applied to the neighboring rings outward and inward.

This method, called the tilted-ring method, is effective for highly inclined galaxies with large $i$. However, the functional shape becomes less sensitive to $i$ in face-on galaxies. Begeman (1989) extensively studied the tilted-ring method, and concluded that it is difficult to determine inclinations for galaxies that are less inclined than $40\deg$.

\subsection{Determination of $\vrot$ for given $i$}

If the inclination angle $i$ is given by another method, a convenient way to derive a rotation curve is to simply measure radial velocities along the major axis. The result is not sensitive to position angle of the major axis according to the above equation's weakly dependency on $\theta$ around 0. Given the inclination $i$, the rotation velocity is obtained by
\begin{equation}
V_{\rm rot}={\vr(r,0) /  \sin i}.
\label{eq_vrot}
\end{equation}
Inclination angle is often obtained from the major-to-minor axial ratio of isophote contour ellipses on optical images. An alternative way is to compare the integrated HI line width with that expected from the Tully-Fisher relation (Shetty et al. 2007).   

However, as equation \ref{eq_vrot} trivially shows, the error of obtained rotation velocity is large for small $i$, and the result even diverges for a face-on galaxy with $i\sim 0\deg$. 

\subsection{Determination of $i$ for given $\vrot$}  

Equation \ref{eq_vrot} is rewritten as
$
\sin i= \vr(r,0) / \Vrot,
$
which means that the inclination can be determined by measuring $\vr(r,0)$, if $\Vrot$ is given. This principle is used in determination of inclination using the Tully-Fisher relation, where one estimates an intrinsic line width using the disk luminosity, and compares it with observed line width to get inclination angle. Shetty et al. (2007) obtained $i=24\deg$ for M51 using this method. 

The above equation can also be applied to individual annulus rings, if the rotation curve is assumed.  It is obvious that the accuracy of determination of $i$ is higher for more face-on galaxies. This method was indeed applied to measure the inclination of the outer HI disk of the face-on galaxy NGC 628 (Kamphuis and Briggs 1992). 

\section{Determination of Inclination for an Assumed Rotation Velocity of M51} 

\begin{figure}
\begin{center} 
\includegraphics[width=7cm]{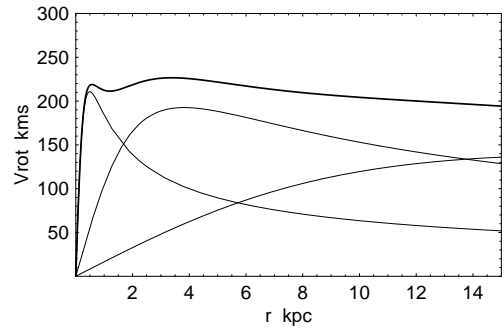}   
\end{center}
\caption{Assumed rotation curve represented by the MN potential for a non-warped disk, which approximately fits the observation within radius 8 kpc of M51.} 
\label{rcmodel}
\end{figure}

We apply the third application described in the previous section to M51 to obtain the radial variation of inclination. Following Kuno and Nakai (1997), we define a universal rotation curve inside 15 kpc radius by Miyamoto-Nagai's (1975) model, and flat rotation at 200 \kms beyond 15 kpc.  Table 1 lists the adopted potential parameters for M51, which approximately reproduce the inner rotation curve at $R<8$ kpc for a fixed inclination angle of $i=24\deg$. Figure \ref{rcmodel} shows the adopted rotation curve with these parameters.

\begin{table} 
\bc
\caption{Miyamoto-Nagai Model for M51} 
\begin{tabular}{lll}  
\hline\hline    
Component & $a_i$ (kpc) & $M_{s}(\Msun)$  \\ 
\hline  
Bulge & 0.35 & $9.33\times 10^{9}$ \\
Disk & 2.7 & $6.02\times  10^{10}$ \\
Halo & 13.45 & $1.55\times  10^{11}$ \\   
  \hline 
  \end{tabular}
  \ec
  \label{tabMN}
\end{table} 

Although the nodal position angle weekly affects the result, we adopt the values obtained for the main disk by Shetty et al (2007), and a fixed value at $162\deg$ beyond 8 kpc. We, then, calculate the best-fit inclination angle $i$ at each radius with interval of 0.05 kpc. 

\begin{figure}
\begin{center} 
\includegraphics[width=7cm]{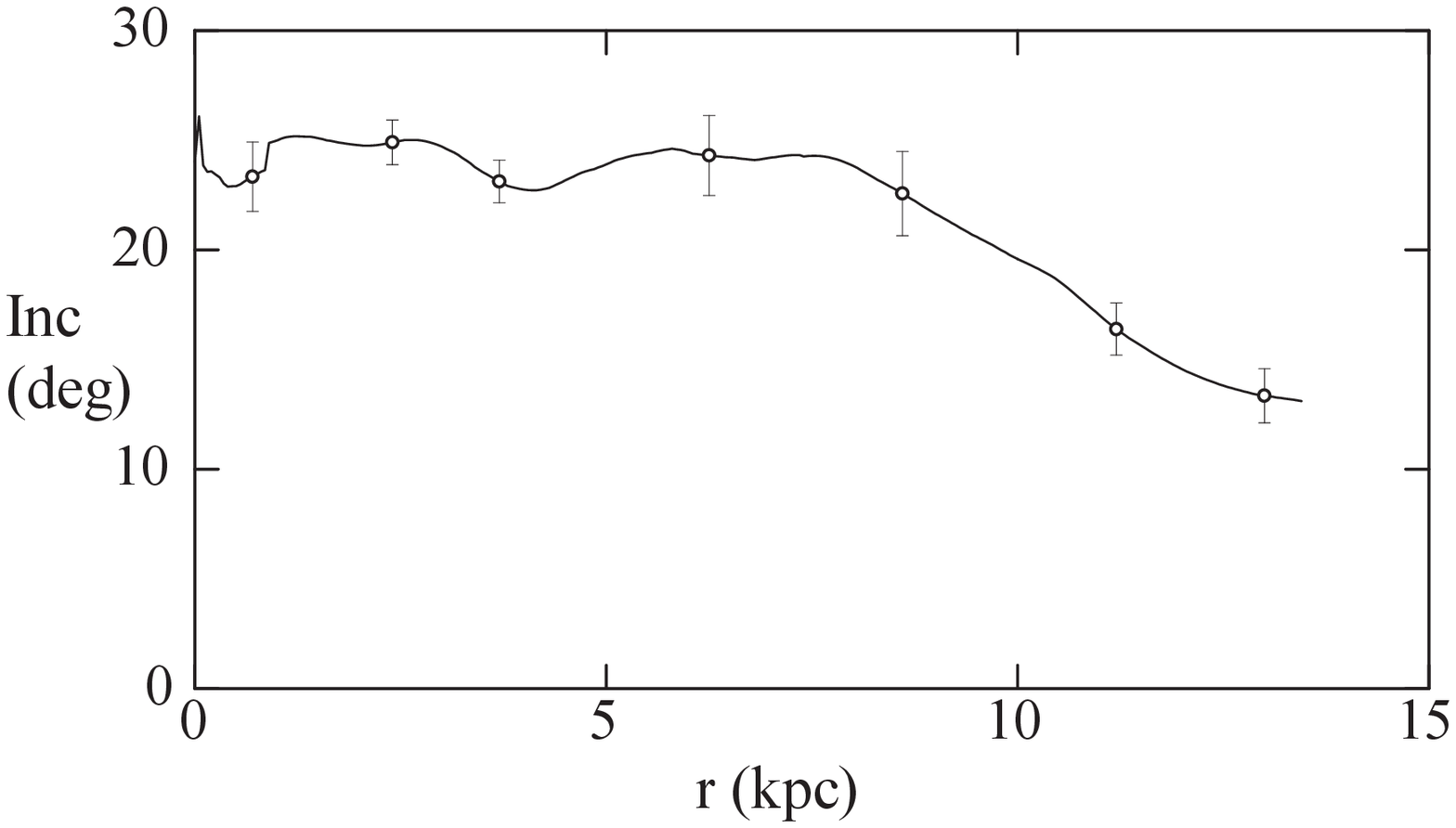}   
\end{center}
\caption{Variation of inclination angle of M51's disk, showing sudden bend at $R=7.5$ kpc.} 
\label{inc}
\begin{center}
\includegraphics[width=7cm]{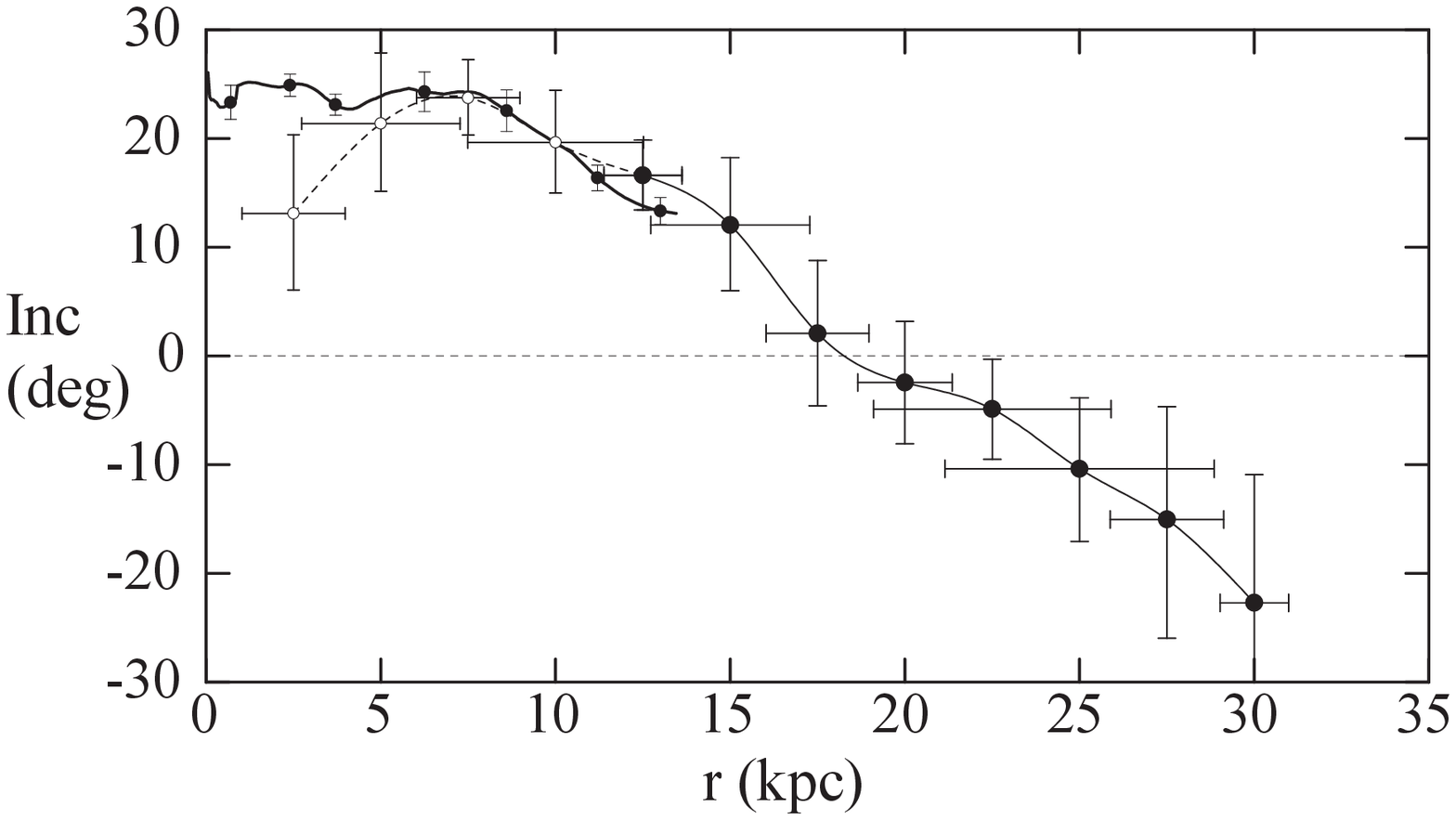}   
\end{center}
\caption{Inclination angle in the outer HI disk. The disk becomes perfectly face-on ($i=0 \deg$) at $R=18$ kpc.} 
\label{rotsInc}
\end{figure}

Figure \ref{inc} shows the thus calculated variation of inclination angle as a function of radius. M51's disk is nearly flat in the inner disk at $r\le 7$ kpc  at $ i \simeq 24\deg$ corresponding to the flat apparent rotation curve. The disk, then, bends suddenly at 7.5 kpc, reaching to inclination angle $i\simeq 12\deg$ at $r=13$ kpc. The result is weakly dependent on the adopted intrinsic rotation curve, in so far as it is usual model. Instead of the MN model, we may assume a flat rotation, which is $\pm 10$ \kms different from the MN model in the analyzed region of M51. This will change in the resulting inclination values by about a few \% . 

Figure \ref{rotsInc} shows the same, but including the outer HI disk. The warping angle reverses at $r\simeq 18$ kpc, where the galaxy becomes perfectly face on. Beyond this radius, the disk is inversely warped, in an opposite sense to the warp of the inner disk. This yielded the apparent counter rotation of the outermost HI disk.
Note, however, the plot at $r>20$ kpc has a larger uncertainty corresponding to that for used rotation curve in figure \ref{rotsAv}.

We comment on the limitation of the accuracy of the outer inclination analysis. The velocity field by Rots et al (1990) shows that the outer HI disk is not symmetric around the galactic center. Particularly, velocities around the companion galaxy NGC 5195 are systematically larger than M51's velocities, and were not used in the present analysis. Nevertheless, the plotted inclination seems to vary smoothly, and the northern and southern values are in agreement with each other.

\section{Discussion}

We have applied a generalized tilted-ring method to derive an extensive warping structure of M51 by assuming a universal rotation curve, which may be called the inverse tilted-ring method, a particular method to face-on galaxies. 
 
\subsection{Counter-warping structure of M51} 
\begin{figure}
\begin{center} 
\vskip 10mm 
\includegraphics[width=8cm]{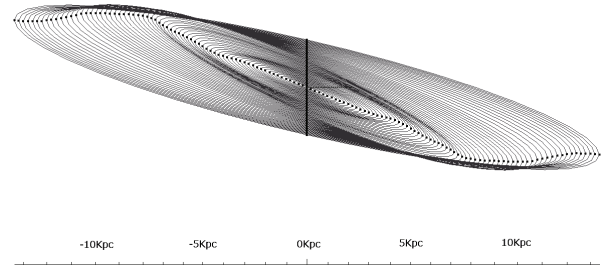}      \\
\includegraphics[width=9cm]{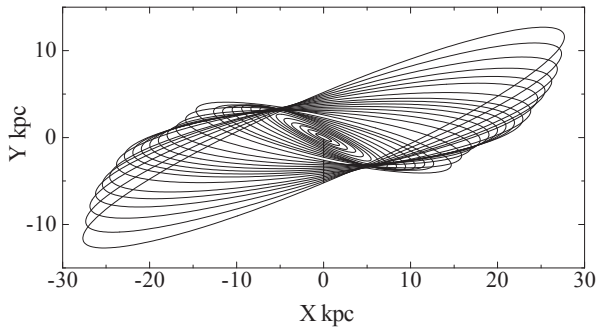}  
\end{center}
\caption{[Top] Bird's eye view of tilted rings for the major disk of M51 seen from 10$\deg$ altitude above the southern major axis. [Bottom] Same, but including HI outskirts drawn every 1 kpc by interpolating obtained inclination values. Note that rings at $r>20$ kpc are a guess from less accurate data, where the HI disk is observed only partially.} 
\label{3d}
\end{figure}

Figure \ref{3d} shows a bird's-eye view of the warped disk of M51 calculated for the obtained inclinations in the present analysis shown in figures \ref{inc} and \ref{rotsInc}, as seen above from the southern major axis at an altitude $10\deg$. The inner warping is consistent with a model drawn by Shetty et al (2007), while the present figure is more quantitative. It is notable that the bending occurs suddenly at $r\sim 7$ kpc, as if the galaxy was broken at this radius. 
This onset radius of warping corresponds to 0.3 Holmberg radii. which is exceptionally smaller than would be expected from Brigg's (1990) rules. On the other hand, leading nodal line derived by Shetty et al (2007) seems consistent with the rule.  
 
The lower panel of figure \ref{3d} shows the outermost HI rings. Apparently counter-rotating HI disk, as observed by Appleton et al. (1986) and Rots et al. (1990), is naturally understood by a counter warping disk. 
Inclination of the outermost ring's reaches as large as $i \sim -20\deg$, or the outermost disk is tilted by $\sim 44\deg$ from the inner main disk, whose inclination is $i=+24\deg$. 
Accordingly, the nearest and farthest parts of rings are overlapping on the line of sight. Note, however, that rings at $r> \sim 20$ kpc are a guess from uncertain outermost rotation curve, and the minor axis regions are not detected in the current HI observations.  

\subsection{Tidal penetration}
M51 is  known for its tidal interaction with the companion. The here obtained warping structure agrees well with a numerical simulation of tidal interaction of M51 and NGC 5195 by Dobbs et al. (2011). Both the simulation and observation shows that the disk is bent suddenly at  $r\sim 7$ kpc. The observed bending angle, $27\deg$, the angle between the inner and outer disks, agrees with the simulated bending angle, $30\deg$. Hence, the present analysis observationally supports the tidal penetration model of the companion through the main disk at $r\sim 7$ kpc about 120 Myr ago (Dobbs et al. 2011).

\subsection{Streaming motion}
M51 is known for large non-circular streaming motions (Kuno and Nakai 1997; Shetty et al. 2007). However, the observed CO-line position-velocity diagram within the main molecular disk at $r<7$ kpc looks quite normal, showing $\pm 20$ \kms deviation from smooth curve. Hence, the streaming motion, which will appear as wavy perturbation of rotation velocity, may affect the resulting inclination by $\pm 10$\% or less.

\subsection{Inverse tilted-ring method}
The present method is a particular usage of the tilted-ring method, and is applicable to any face-on galaxies. In fact, we applied it to IC 342, and obtained a flat disk at almost a constant inclination corresponding to its quite normal flat rotation curve. The method was also used by Kamphuis and Briggs (1992) to obtain warping structure in the face-on galaxy NGC 628, where they assumed a constant rotation velocity. 

We may remind that face-on galaxies with $i\le 30\deg$ shares $\sim 13$ \% of total galaxies. In these face-on galaxies, radial velocities yield large uncertainty in their derived rotation velocity. Instead, if their intrinsic rotation curves can be assumed, the data may be useful for geometrical investigations of galactic disks.

{}


\begin{thebibliography}{}

\bibitem[Appleton et al.(1986)]{1986MNRAS.221..393A} Appleton, P.~N., Foster, P.~A., \& Davies, R.~D.\ 1986, \mnras, 221, 393  

\bibitem[Begeman(1989)]{1989A&A...223...47B} Begeman, K.~G.\ 1989, \aap, 223, 47  
 

\bibitem[Bosma(1981)]{1981AJ.....86.1791B} Bosma, A.\ 1981, \aj, 86, 1791 
 
\bibitem[Briggs(1990)]{1990ApJ...352...15B} Briggs, F.~H.\ 1990, \apj, 352, 
15  

\bibitem[Dobbs et al.(2010)]{2010MNRAS.403..625D} Dobbs, C.~L., Thesis, C., Pringle, J.~E., \& Bate, M.~R.\ 2010, \mnras, 403, 625  

\bibitem[Garcia-Burillo et al.(1993)]{1993A&A...274..123G} Garcia-Burillo, S., Guelin, M., \& Cernicharo, J.\ 1993, \aap, 274, 123 
 
\bibitem[J{\'o}zsa et al.(2007)]{2007A&A...468..731J} J{\'o}zsa, G.~I.~G., Kenn, F., Klein, U., \& Oosterloo, T.~A.\ 2007, \aap, 468, 731 
 

\bibitem[Kamphuis \& Briggs(1992)]{1992A&A...253..335K} Kamphuis, J., \& Briggs, F.\ 1992, \aap, 253, 335 

\bibitem[Koda et al.(2011)]{2011ApJS..193...19K} Koda, J., Sawada, T., Wright, M.~C.~H., et al.\ 2011, \apjs, 193, 19 

\bibitem[Haynes et al.(1978)]{1978AJ.....83..938H} Haynes, M.~P., Giovanelli, R., \& Burkhead, M.~S.\ 1978, \aj, 83, 938 

\bibitem[]{} Miyamoto, M.,  Nagai, R. 1975, PASJ, { 27}, 533.  

\bibitem[McGaugh et al.(2000)]{2000ApJ...533L..99M} McGaugh, S.~S., 
Schombert, J.~M., Bothun, G.~D., 
\& de Blok, W.~J.~G.\ 2000, \apjl, 533, L99 


\bibitem[Kuno \& Nakai(1997)]{1997PASJ...49..279K} Kuno, N., \& Nakai, N.\ 1997, \pasj, 49, 279 


\bibitem[Kuno et al.(1995)]{1995PASJ...47..745K} Kuno, N., Nakai, N., Handa, T., \& Sofue, Y.\ 1995, \pasj, 47, 745 


\bibitem[Nakai et al.(1994)]{1994PASJ...46..527N} Nakai, N., Kuno, N., 
Handa, T., \& Sofue, Y.\ 1994, \pasj, 46, 527 


\bibitem[Newton(1980)]{1980MNRAS.191..169N} Newton, K.\ 1980, \mnras, 191, 
169  

\bibitem[Persic et al.(1996)]{1996MNRAS.281...27P} Persic, M., Salucci, P., 
\& Stel, F.\ 1996, \mnras, 281, 27 

 

\bibitem[Rand(1993)]{1993ApJ...410...68R} Rand, R.~J.\ 1993, \apj, 410, 68 
 
\bibitem[Roberts \& Warren(1970)]{1970A&A.....6..165R} Roberts, M.~S., \& Warren, J.~L.\ 1970, \aap, 6, 165 

\bibitem[Rogstad et al.(1974)]{1974ApJ...193..309R} Rogstad, D.~H., 
Lockhart, I.~A., \& Wright, M.~C.~H.\ 1974, \apj, 193, 309 



\bibitem[Rots et al.(1990)]{1990AJ....100..387R} Rots, A.~H., Bosma, A., van der Hulst, J.~M., Athanassoula, E., \& Crane, P.~C.\ 1990, \aj, 100, 387  

\bibitem[Rubin et al.(1980)]{1980ApJ...238..471R} Rubin, V.~C., Ford, W.~K.~J., \& .~Thonnard, N.\ 1980, \apj, 238, 471 
 
\bibitem[Salucci et al.(2007)]{2007MNRAS.378...41S} Salucci, P., Lapi, A., Tonini, C., et al.\ 2007, \mnras, 378, 41 

\bibitem[Sancisi(1976)]{1976A&A....53..159S} Sancisi, R.\ 1976, \aap, 53, 159 
 
 
\bibitem[Shetty et al.(2007)]{2007ApJ...665.1138S} Shetty, R., Vogel, S.~N., Ostriker, E.~C., \& Teuben, P.~J.\ 2007, \apj, 665, 1138  
  
\bibitem[]{}Sofue, Y., 1996, ApJ 458, 120  

\bibitem[Sofue(1997)]{1997PASJ...49...17S} Sofue, Y.\ 1997, \pasj, 49, 17 



\bibitem[Sofue(1998)]{1998PASJ...50..227S} Sofue, Y.\ 1998, \pasj, 50, 227

 \bibitem[Sofue et al.(1999)]{1999ApJ...523..136S} Sofue, Y., Tutui, Y., 
Honma, M., et al.\ 1999, \apj, 523, 136 



\bibitem[]{}Sofue, Y., Rubin, V.C. 2001 ARAA 39, 137  
 

\bibitem[Tilanus \& Allen(1991)]{1991A&A...244....8T} Tilanus, R.~P.~J., \& Allen, R.~J.\ 1991, \aap, 244, 8 
 

\bibitem[Tully(1974)]{1974ApJS...27..437T} Tully, R.~B.\ 1974, \apjs, 27, 437 
 

\end{thebibliography}
\end{document}